# Atom identification in bilayer moiré materials with Gomb-Net

Austin C. Houston[1], Sumner B. Harris[2*], Hao Wang[3], Yu-Chuan Lin[4], David B. Geohegan[1], Kai Xiao[2], Gerd Duscher[1*]

[1] Department of Materials Science and Engineering, University of Tennessee, Knoxville, TN 37996
[2] Center for Nanophase Materials Sciences, Oak Ridge National Laboratory, Oak Ridge, TN 37831
[3] Min H. Kao Department of Electrical Engineering and Computer Science, The University of Tennessee, Knoxville, Tennessee 37916, United States
[4] Department of Materials Science and Engineering, National Yang Ming Chiao Tung University, Hsinchu City, 300, Taiwan

*Correspondence should be addressed to: harrissb@ornl.gov or gduscher@utk.edu

**Abstract**

Moiré patterns in van der Waals bilayer materials complicate the analysis of atomic-resolution images, hindering the atomic-scale insight typically attainable with scanning transmission electron microscopy. Here, we report a method to detect the positions and identities of atoms in each of the individual layers that compose twisted bilayer heterostructures. We developed a deep learning model, Gomb-Net, which identifies the coordinates and atomic species in each layer, deconvoluting the moiré pattern. This enables layer-specific mapping of quantities like strain and dopant distributions, unlike other commonly used segmentation models which struggle with moiré-induced complexity. Using this approach, we explored the Se atom substitutional site distribution in a twisted fractional Janus $WS_2$-$WS_{2(1-x)}Se_{2x}$ heterostructure and found that layer-specific implantation sites are unaffected by the moiré pattern's local energetic or electronic modulation. This advancement enables atom identification within material regimes where it was not possible before, opening new insights into previously inaccessible material physics.

Over the past decade, twisted van der Waals bilayers (TvdWBs) have drawn significant attention for their unique and often tunable properties, their promise for creating elusive states of matter, and for studying the strongly correlated physics that emerges[1]. The moiré lattice created by two monolayers of graphene or transition metal dichalcogenides (TMDs) modulates electronic and topological properties on the nanoscale[2], producing phenomena from perfect arrays of quantum emitters to excitonic superlattices with tunable, giant spin-orbit coupling[3]. In bilayer graphene, unconventional superconductivity emerges at a ~1.1° "magic twist angle", where the Fermi velocity drops to zero and energy bands become flat[4].

Each atom in a monolayer of these types of materials is readily imaged with high-angle annular dark-field scanning transmission electron microscopy (HAADF-STEM), and is identified by the image contrast, which is roughly proportional to the square of the atomic number Z. Thus, any defect or distortion from the ideal crystal lattice is also directly identifiable. Microscopists have developed numerous traditional techniques for identifying atom positions and species over the past half century. However, when these methods fail, deep learning models such as AtomAI[5] and AtomSegNet[6] are employed for atom finding, defect identification, strain mapping, and local structure analysis[7–12]. Neural networks like these are semantic segmentation models based on the U-Net[13,14] architecture, which classify image pixels into categories. Such segmentation models offer advantages over traditional image analysis methods, including improved accuracy in atom localization, with minimal parameter tuning and the ability to generate rapid predictions in milliseconds[6].

However, the moiré interference in HAADF-STEM images of TvdWBs introduces variations in the expected Z-contrast for each atomic species, caused by differing degrees of positional overlap between atoms in the two layers. This gradual change in atomic stacking through the superlattice prevents reliable atom identification with standard semantic segmentation models. Although Fourier filtering can estimate the repeating atomic lattices, it discards information about inhomogeneities and atomic species. This insensitivity to the local environment is particularly significant because strain and defect engineering have a significant effect on TvdWBs properties[15]. While defects in 2D materials are often detrimental, some novel optical properties are enabled by their presence[16,17], such as point defects in $WSe_2$ monolayers acting as single-photon emitters[18]. Clearly, the location, quantity, and local structure of point defects in these systems determine their properties. Although multi-slice ptychography can identify atoms in moiré materials, it is computationally intensive, requires extensive parameter optimization, expensive equipment, and significant expertise. Identification of atomic species in TvdWBs from HAADF-STEM images therefore remains a challenge that must be addressed to gain a deeper understanding of how the local atomic structure of moiré materials affects their properties.

Here, we develop a deep learning model called Gomb-Net (a multi-branch U-Net using a groupwise combinatorial loss) that can identify atoms in each individual layer of TvdWBs, enabling us to probe the local atomic structure of these moiré materials. Gomb-Net uses a multi-branch decoder U-Net architecture to perform image segmentation on HAADF-STEM images, effectively deconvoluting the individual monolayers from the moiré pattern to reveal layer-wise distributions of dopant atoms, defects, or strain. We demonstrate improved performance compared to the conventional U-Net architecture for this task using images from two TvdWB test cases: twisted bilayer graphene (TBG) and a twisted TMD heterostructure of $WS_2$-$WS_{2(1-x)}Se_{2x}$. Gomb-Net can be trained in minutes on modern personal computers and can generate predictions in milliseconds, making advanced analysis of moiré matter widely accessible and ideal for real-time atom identification and analysis during automated or autonomous STEM experimentation.

The TBG system, being one of the simplest yet most impactful moiré systems, serves as the ideal test case for evaluating the capabilities of Gomb-Net against other atom finding techniques. A simulated HAADF-STEM image of TBG at atomic resolution (**Fig 1a**) shows repeating moiré unit cells, where atom identification seems feasible at the highly symmetric AA stacking centers. However, moving away from the clearly defined carbon rings into intermediate stacking regions, the moiré pattern complicates the interpretation of the Z-contrast - for both the human eye and traditional atom-finding methods. Thus, we developed Gomb-Net to accurately identify the positions of atoms in each individual layer, thereby deconvoluting the moiré pattern for atomic-scale analysis.

Gomb-Net achieves this through two modifications to the standard U-Net model. First, we employ a physics-informed loss function during training, which we call the groupwise combinatorial loss (Gomb-Loss, detailed in the Methods section of the Supporting Information). This loss function disregards the layer ordering (physically, which layer is on "top" of the other does not affect HAADF-STEM image formation), in favor of the sorting the atoms into individual layers. Mathematically, Gomb-Loss prioritizes interlayer coherence and accounts for multiple stacking order scenarios through a series of grouping outputs and averaging reciprocals, similar to a harmonic mean. We train Gomb-Net on simulated HAADF-STEM images of the TvdWB system of interest, TBG in this first example. Accordingly, Gomb-Net's performance and limitations are critically influenced by the quality and diversity of the simulated datasets used for training as well as their resemblance to real-world datasets. Detailed descriptions of the training data generation are included in the Methods section of the Supporting Information.

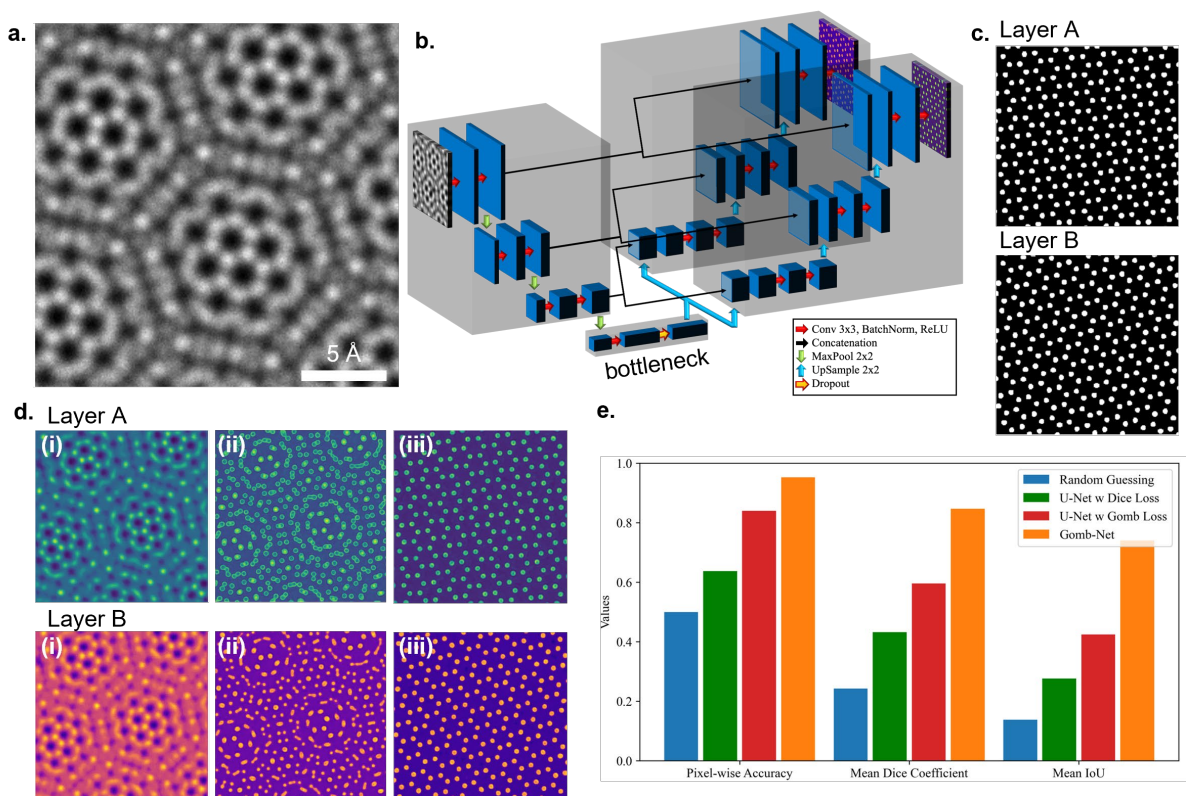

**Figure 1. Atom-finding in Moiré lattices is enabled by Gomb-Net. a,** Simulated HAADF-STEM test image of atomically resolved twisted bilayer graphene, used as network input (not seen during training). **b,** Schematic of the network architecture with a single encoder before the bottleneck layer, and two separate decoders after the bottleneck layer. Skip connections (black) help preserve fine detail. **c,** Binary images showing the predicted carbon atom positions for each layer derived from the input. **d,** Raw network outputs from the test image for each layer using (i) U-Net, (ii) U-Net with the groupwise combinatorial loss function, and (iii) Gomb-Net. **e,** Performance comparison of the different networks on a test dataset of 800 images.

The second technical development is the adoption of a multi-branch U-Net architecture, shown in **Fig 1b**, which allows for multiple decoders. Each branch is responsible for segmenting a single atomic layer (the two decoders in this example are visible in **Fig 1b, right**). This dual-decoder architecture has been used for multi-task learning on biological data[19] and we adapt it here for segmentation of HAADF-STEM images of TvdWBs. After training, Gomb-Net produces pixel-
5

wise classifications for atom and layer identity in the bilayer heterostructure (**Fig 1c**), facilitating the analysis of strain, defects, and composition in TvdWBs.

To show the necessity of these technical developments, **Fig 1d** compares the raw outputs of three networks: (i) standard U-Net trained with dice loss, (ii) U-Net trained with Gomb-loss, (iii) and Gomb-Net trained with Gomb-loss. The standard U-Net classifies predominantly based on pixel brightness and introducing Gomb-loss localizes the atom predictions. However, both networks output nearly identical predictions for each layer. In contrast, the multi-branch architecture of Gomb-Net leads to branch specialization, allowing individual atoms to be separated by layer. To evaluate network performance, we compared accuracy metrics from these three networks to random guessing using a test dataset of 800 simulated images (**Fig 1e**). Gomb-Net achieves pixel-wise accuracy of 0.98 on the test dataset while the U-Nets reach 0.86 maximum. The mean Intersection over Union (IOU) highlights Gomb-Net's performance in minimizing false positives and negatives with 0.74 IOU, compared to 0.39 maximum for the U-Nets. Accurately identifying atom positions with minimal false positives and false negatives is essential, as incorrect predictions significantly impact the structural analysis of materials.

Next, we demonstrate the performance of Gomb-Net on real HAADF-STEM images of twisted bilayer graphene. **Fig 2a** shows an experimental image of a bilayer graphene sample with a ~ 26 degree twist, exhibiting a moiré pattern which makes identifying locations of individual C atoms challenging via traditional methods. This image is passed to the pre-trained Gomb-Net for segmentation and layer classification, the raw model outputs are shown in **Fig 2b**. A threshold of 0 is applied to extract gated outputs (**Fig 2c**) which are then translated into atomic coordinates (x, y positions) via blob finding and computing the center-of-mass of each blob (**Fig 2d**), providing sub-pixel accuracy. To validate the network predictions, we calculate the C-C distance for all the identified atoms in both layers and compare to the known ideal C-C bond distance in bilayer graphene (1.42 Å)[20]. **Fig 2e** shows a histogram of predicted C-C distances with the mean value of 1.39 Å and full width half max (FWHM) 0.12 Å, which gives a 2.11 % difference from the anticipated value and is well within the experimental uncertainty, especially considering that TBG exhibits heterostrain over 0.2 % across different regions of the moiré[21]. We achieve near pixel-level precision in this measurement given that the pixel size is ~ 0.11 Å/pixel.

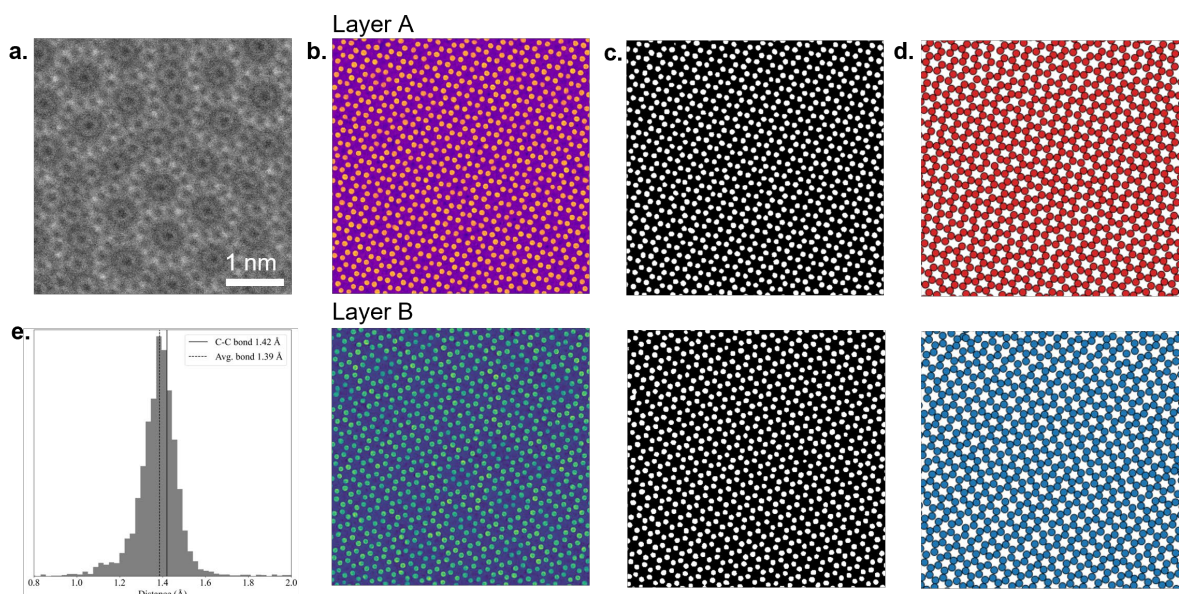

**Figure 2. Atom-finding in twisted bilayer graphene provides pixel-level accuracy for carbon atom positions in each layer. a,** Experimental atomically resolved HAADF-STEM image of twisted bilayer graphene. **b,** Raw Gomb-Net outputs for layer A (top row) and layer B (bottom row) **c,** Binarization of the outputs with gating applied at threshold of zero to segment carbon atoms. **d,** Atom positions of both graphene layers, extracted from the binary images. **e,** Distribution of C-C bond distances for validation to a known physical value.

To further demonstrate our approach, we apply Gomb-Net to a system where the moiré pattern is complicated by both twist angle and a difference in lattice constant: bilayer $WS_2$-$WS_{2(1-x)}Se_{2x}$ Janus TMDs. Specifically, we examine a twisted $WS_2$-$WS_2$ bilayer where the Se atoms have been preferentially substituted into the S sites on the outermost chalcogen sublayer to form a fractional Janus alloy[22,23]. Janus TMDs are interesting due to their intrinsic asymmetry which leads to a built-in electronic dipole and offers optical and catalytic properties that differ from the parent

TMDs[24] and are particularly promising for producing tunable moiré matter to control interlayer excitons[25]. The modulation of their properties by moiré patterns is largely unknown. Moreover, questions remain about how strain or and local structure may influence the doping or substitutional site preferences during the synthesis process[26].

We synthesized the fractional Janus $WS_2$-$WS_{2(1-x)}Se_{2x}$ bilayer sample using pulsed laser deposition (PLD). In this process, the $WS_2$ bilayer is exposed to the PLD plasma plume produced by ablating a solid Se target. The plume is composed of Se atoms and molecular clusters with a maximum kinetic energy controlled to ~ 8.5 eV/atom (see Methods in the Supporting Information). Previous experiments with $WS_2$ monolayers, using in situ Raman spectroscopy and ex situ STEM imaging, found that Se atoms are initially implanted primarily in the topmost chalcogen sublayer of $WS_2$. With a monolayer, this transformation proceeds predominantly in a layer-by-layer fashion until the $WS_2$ monolayer is fully converted to $WSe_2$[23]. A fractional Janus layer is formed at stoichiometry x < 0.5 and a full Janus WSSe layer is formed at stoichiometry x = 0.5[23].

Gomb-Net is necessary to answer the following question: does the modulated local coupling of the TMD layers induced by the moiré pattern significantly influence the implantation site probability of Se into the exposed TMD layer? This relationship can be elucidated from HAADF-STEM imaging experiments of these twisted $WS_2$-$WS_{2(1-x)}Se_{2x}$ structures, but analysis is significantly more complicated than in the bilayer graphene case. Graphene segmentation required prediction of only 2 classes ($C_A$ and $C_B$), while this case requires 6 classes: W atoms, S-S columns, and Se-S columns in both monolayers (labeled A and B). If Gomb-Net can identify atoms and defects in this system, it would provide a better understanding of the mechanisms of Se implantation and the effects of local defects and strain on the structure and properties of twisted TMD alloy bilayers.

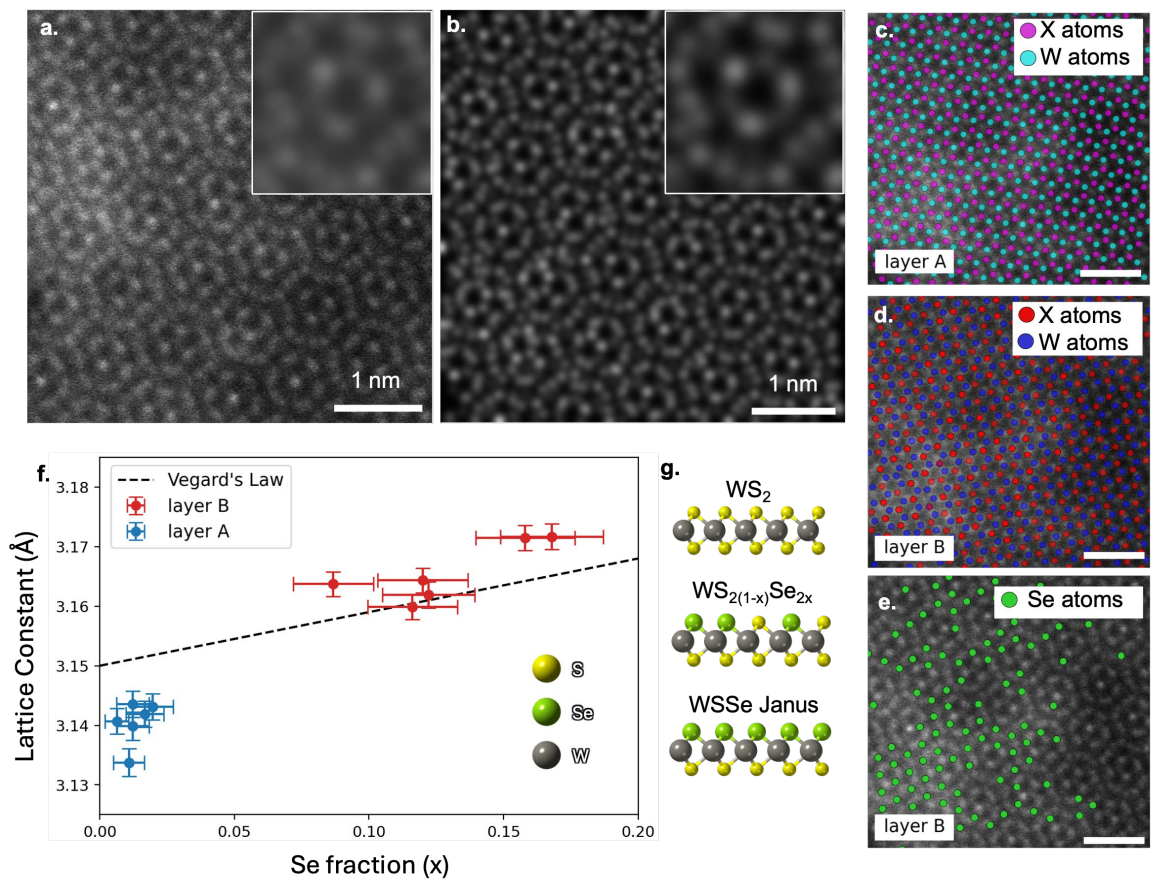

**Figure 3. Atom-finding in twisted $WS_2$-$WS_{2(1-x)}Se_{2x}$ bilayers suggests random, layer-selective Se substitution. a,** Experimental, atomically resolved HAADF-STEM image of the twisted $WS_2$-$WS_{2(1-x)}Se_{2x}$ heterostructure input to Gomb-Net. **b,** Simulated reconstruction image from Gomb-Net atomic position predictions. Insets: magnified regions for comparison. **c,** Tungsten and chalcogen (X) atom positions found in layer A. **d,** Tungsten and chalcogen atom positions found in layer B. **e,** Distribution of all Se atoms in layer B. **f,** Lattice constant as a function of stoichiometry, calculated from the Gomb-Net outputs of 6 images. **g,** Atomic models of $WS_{2(1-x)}Se_{2x}$ components for Vegard's law calculation in **f**.

Gomb-Net was trained on simulated data for this system and used to analyze an experimental HAADF-STEM image, shown in **Fig 3a**. The model extracts two sets of atomic positions and types, which are used to reconstruct a simulated HAADF-STEM image (**Fig 3b**). The reconstructed image serves to visually confirm network performance on experimental data, where the ground truth is not known. This reconstructed image agrees remarkably well with the raw image, as shown in the insets featuring a high-symmetry region in the moiré lattice. Another region is shown in **Fig 3c-e**, where Se-S columns are present in layer B (**Fig 3d-e**) but absent in the layer A (**Fig 3c**). Across six similar regions, 98 % of all Se atoms (identified by the presence of Se-S columns) were found on the $WS_2$ layer exposed to the Se PLD plume. This is consistent with the previous experimental observations[23]. The atomic coordinates derived from model predictions for the six cropped regions are used to measure the lattice constant and stoichiometry of the individual layers **Fig 3f**. We can compare this ratio to Vegard's law, which states that the lattice parameter of a system with two constituents is approximately a weighted sum of the two constituents' lattice parameters. Notably, the lattice parameter of layer A corresponds to ~ 2 % Se substitution, which agrees with the direct atom counting above.

With Se positions known, determination of whether the Se is preferentially substituted in the moiré lattice depends on labeling the different moiré sites. We approach this by first calculating the order parameter vector[27], $u^{A/B}$, which represents the x-y projection distance between atoms in layer A and layer B. For each pixel in the image (**Fig 4a**), the nearest tungsten (W) and chalcogen columns (X) are identified, and the distance between them is calculated. Maps of the order parameters $u^{W/X}$ and $u^{X/W}$ are shown in **Fig 4b-c**. To produce a map of moiré stacking, we take the Euclidean norm of $u^{W/X}$ and $u^{X/W}$. The resulting local stacking order map (**Fig 4c**), or moiré map, shows that a region with perfect AB stacking has a value of 0, a region with perfect AA' stacking corresponds to the lattice parameter, *d*, and a region with perfect A'B stacking corresponds to $d\sqrt{2}$ (**Fig 4d**). Since the lattice parameter changes with Se concentration **(Fig 3f),** we use the average value of 3.15 Å.

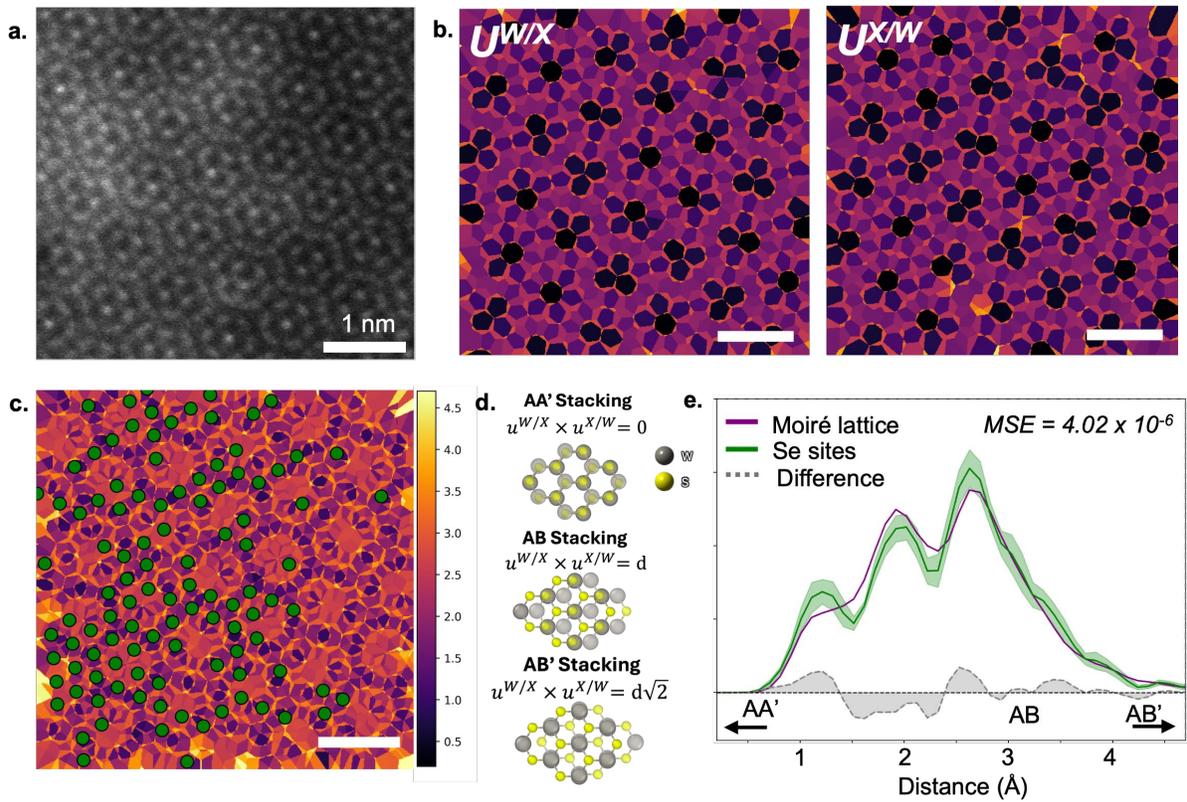

**Figure 4. Mapping the Se site distribution shows no preferential substitutional sites related to the moiré lattice. a,** Experimental, atomically resolved HAADF-STEM image of a twisted $WS_2$-$WS_{2(1-x)}Se_{2x}$ heterostructure. **b,** Order parameter maps calculated from Gomb-Net atomic position predictions. Left: $U^{W/X}$, the in-plane distance from layer A tungsten to the nearest layer B chalcogen column. Right: $U^{X/W}$, the in-plane distance from layer A chalcogen column to the nearest layer B tungsten. **c,** Local stacking order, calculated by $U^{W/X} \times U^{X/W}$, with Se-S columns (green). **d,** Schematic of local stacking order configurations and their respective values in terms of lattice parameter, $d$. **e,** Distributions of local stacking order in the moiré lattice (purple) and Se site occupancy (green) shows no preferential sites for Se substitution.

The final step to determine preferential occupation of the moiré lattice is to associate each Se-S column with the nearest moiré map value. This is done for six cropped regions of the original HAADF-STEM image, and the resulting distributions are shown in **Fig 4e**. To estimate the probability density function of this distribution, we used Kernel Density Estimation (KDE) with a Gaussian kernel. The KDE of the 1.3 million moiré sites is shown in purple, while the KDE of the 333 Se-S column positions found across the images is shown in green. The mean KDE of each distribution is plotted as a solid line, with plus/minus one standard deviation shown as the shaded region in **Fig 4e**. Notably, the standard deviation of the moiré distance KDE is too small to be visible at this scale. The mean squared error between these two density-normalized distributions is $4.02 \times 10^{-6}$, indicating no preferential occupation of Se in the moiré lattice.

This result was unexpected, as we hypothesized the barrier to implantation would be higher where the underlying lattice was more neatly stacked. We expected to see more Se-atoms implanted in the regions between perfect stacking orders. From these results, two scenarios seem likely. In the first scenario, the difference in energy threshold for implantation across the moiré is small compared to the energy of the incident Se plume and, thus, the Se atoms implant randomly. In the alternative scenario, the Se do implant preferentially on the moiré but diffuse randomly across the sample – the implantation process is known to create point defects which would facilitate Se diffusion, especially at elevated temperatures. In either scenario, there is no final preference for Se occupation on the moiré lattice.

Another finding is that a fraction of the chalcogen sites on the bottom layer (~ 2 % in this case) are implanted with Se atoms before the primary exposed layer is fully converted to Janus WSSe or $WSe_2$. When the model predicts on simulated images from the test dataset for cases where Se atoms are known to exist in only one layer, it falsely predicts Se atoms in the other layer with an error rate on the order of 0.5 %. Thus, it is likely that some of these Se atoms are false positives, while others are real. Regardless, these errors are still low enough to make statistically accurate statements about implantation as a function of local order; however, it suggests that a minority fraction of Se atoms penetrate through the top monolayer. The analysis enabled by Gomb-Net opens opportunities to study the full range of implantation conditions and twist angles to reveal the mechanisms of Se implantation into bilayer TMDs by PLD.

Gomb-Net was built with the intention that scientists working with atomically resolved images of moiré materials can adapt the methodology to their own problems. The code is modular and flexible, with full control of hidden layers, loss function, output layers, and other hyperparameters, and was made to be adaptable to any moiré problem. The case of twisted $WS_2$-$WS_{2(1-x)}Se_{2x}$ bilayer atom identification requires 6 output classes and the network performs well. Cases with > 6 output classes or additional layers, such as a three-layer moiré material, are feasible by incorporating more decoder branches and adding additional terms to the loss function but have not been assessed yet. The key factor and limitation to Gomb-Net's performance, as with any supervised learning model, is the degree of similarity between the training dataset and the real experimental data. For example, if the goal is to identify lateral grain boundary structures in bilayer stacks of 2D materials, the training data must be constructed with simulated data that includes a variety of anticipated structures, likely determined by density functional theory or molecular dynamics simulations.

To summarize, we have demonstrated the ability to locate the positions, atomic identities, and layer identities of atoms in twisted van der Waals bilayers from HAADF-STEM images using a modified U-Net segmentation model called Gomb-Net. The key enabling modifications are a custom loss function and a multi-branch decoder design (here two decoders, one for each layer) that allows segmentation of each layer as a whole. For twisted bilayer graphene, we achieved high accuracy in distinguishing carbon atoms in the individual layers and used the predicted atomic positions to measure the lattice constant with pixel-scale precision, effectively deconvoluting the moiré pattern. Extending the approach to the more complex case of a fractional Janus $WS_2$-$WS_{2(1-x)}Se_{2x}$ bilayer enabled us to explore possible chalcogen-site substitution preferences related to the moiré lattice. We found that, under the tested sample synthesis conditions, Se atoms occupy the chalcogen sites without preference for specific moiré lattice sites. Overall, these findings highlight Gomb-Net's precision and adaptability for studying a wide range of moiré materials, paving the way for future investigations of atomic-scale phenomena in 2D twisted heterostructures. Additionally, Gomb-Net can be deployed on modern personal computers, making advanced analysis of moiré materials more accessible and ideal for real-time analysis during autonomous STEM experimentation.

**Data availability**

The data that support the findings of this study are openly available at:
https://drive.google.com/drive/folders/1tDF283xry5op3t594oBUlcNLKbjRTV7C?usp=sharing

**Code availability**

The code for this study can be accessed via this link: https://github.com/ahoust17/Gomb-Net.git

**Acknowledgements**

This work was supported by the U.S. Department of Energy, Office of Science, Basic Energy Sciences, Materials Sciences and Engineering Division. Development of machine learning methods was supported by the Center for Nanophase Materials Sciences (CNMS), which is a U.S. Department of Energy, Office of Science User Facility at Oak Ridge National Laboratory. The microscopy and machine learning in this work was partially supported by the AI Tennessee Initiative and the Electron Microscopy Center at the University of Tennessee, Knoxville. For the MOCVD growth of 2D $WS_2$ crystals Y.-C. L. acknowledges funding from NEWLIMITS, a center in nCORE as part of the Semiconductor Research Corporation (SRC) program sponsored by NIST through award number 70NANB17H041.

**Author Declarations**

*Conflict of Interest:*

The authors have no conflicts to disclose.

*Author Contributions:*

**ACH**: Conceptualization (equal); writing—original draft (lead); writing—review & editing (equal); data curation (equal); methodology (equal); software (lead); visualization (lead). Investigation (lead). **SBH**: Conceptualization (equal); writing—original draft (supporting); writing—review & editing (equal); data curation (equal); methodology (equal); software (supporting); Investigation (supporting). **HW**: Investigation (supporting). **Y-CL**: Investigation (supporting). **DGB**: Resources (equal); Writing – review & editing (equal). **KX**: Funding acquisition (lead); Resources (equal); Writing – review & editing (equal). **GD** Resources (equal); Writing – review & editing (equal).

**Supplemental Information for:**

# Atom identification in bilayer moiré materials with Gomb-Net

Austin C. Houston[1], Sumner B. Harris[2*], Hao Wang[3], Yu-Chuan Lin[4], David B. Geohegan[2], Kai Xiao[2], Gerd Duscher[1*]

[1] Department of Materials Science and Engineering, University of Tennessee, Knoxville, TN 37996
[2] Center for Nanophase Materials Sciences, Oak Ridge National Laboratory, Oak Ridge, TN 37831
[3] Min H. Kao Department of Electrical Engineering and Computer Science, The University of Tennessee, Knoxville, Tennessee 37916, United States
[4] Department of Materials Science and Engineering, National Yang Ming Chiao Tung University, Hsinchu City, 300, Taiwan

*Correspondence should be addressed to: harrissb@ornl.gov or gduscher@utk.edu

**Methods**

*Data Generation*

Gomb-Net is trained on separate sets of simulated HAADF-STEM images for the graphene (2000 images) and TMD bilayer (8000 images) cases, split into training, validation, and test subsets with proportions of 70%, 20%, and 10%, respectively. To generate our simulated HAADF-STEM images, we first generate the crystal structure for each bilayer scenario with atomic simulation environment package[1] and use a Gaussian to approximate the scattering potential at each atomic position. This scattering potential is convoluted with a point spread function that represents the ideal case of a well-focused and aberration corrected electron probe, an airy disk, to generate the simulated Z contrast image. We include variations in layer twist angle, crystal rotation, atomic vacancy count, atomic vibration, atomic size, number of holes, hole and edge boundaries, and relative magnification. Poisson noise is randomly added to simulated images to replicate the stochastic nature of electron counting in actual HAADF imaging experiments. Large-scale Gaussian noise is randomly added to approximate intensity variations caused by contamination on the sample. Images for the $WS_2$-$WS_{2(1-x)}Se_{2x}$ dataset are generated from a $WS_2$ bilayer where S atoms are randomly replaced with Se according to one of the following randomly chosen scenarios: pristine (no Se substitution), single chalcogen sub-layer substitution, single crystal layer substitution, or substitution into both layers. This complexity in training data prepares the network for most S-Se substitution scenarios. The full range of varied parameters is shown in **Table S1** and the process is visualized in **Figure S1**. The code used to generate the training data is available online: https://github.com/ahoust17/DataGenSTEM.git.

**Dataset Generation**

| Parameter | Graphene (bilayer) dataset | WSSe dataset |
|---|---|---|
| Total images | 2000 | 8000 |
| Crops per batch | 20 | 10 |
| Image size (pixels) | 256 | 512 |
| Phonon sigma | uniform(0.02, 0.1) | uniform(0.05, 0.15) |
| Rotation layer 1 (r1) | uniform(0, 360) | uniform(0, 360) |
| Rotation layer 2 | uniform(0, 360) | r1+n*normal(mean=15, var=0.5) |
| Atom size | normal(mean=0.175, var=0.01) | normal(mean=0.18, var=0.01) |
| Shot noise | uniform(0.6, 0.9) | uniform(0.88, 0.98) |
| Blob size | 0 | uniform(0.005, 0.01) |
| Blob intensity | 0 | uniform(1, 5) |
| Magnification | uniform(0.2, 0.35) | uniform(0.2, 0.35) |
| Vacancies | 10 | int(normal(mean=18, var=5)) |
| Se conversion | NA | uniform(0,1) |
| Scenario | NA | uniform choice() |

**Table S1.** Parameters used for dataset generation for both datasets in this work. The distribution names (uniform, normal) show that a parameter was chosen at random for each batch of image generation. Only two parameters were chosen independently for every single image, instead of batchwise: shot noise and magnification. Phonon sigma refers to the parameters used for the frozen phonon approximation. Gaussian blobs were applied in the WSSe case because the samples used to gather the experimental data were significantly dirty, leading to blob-like noise in the image - the network predictions should be robust against this kind of noise. Notably, the interlayer rotation of the WSSe dataset was a normal distribution centered around the known experimental twist angle for the two layers, determined by FFT.

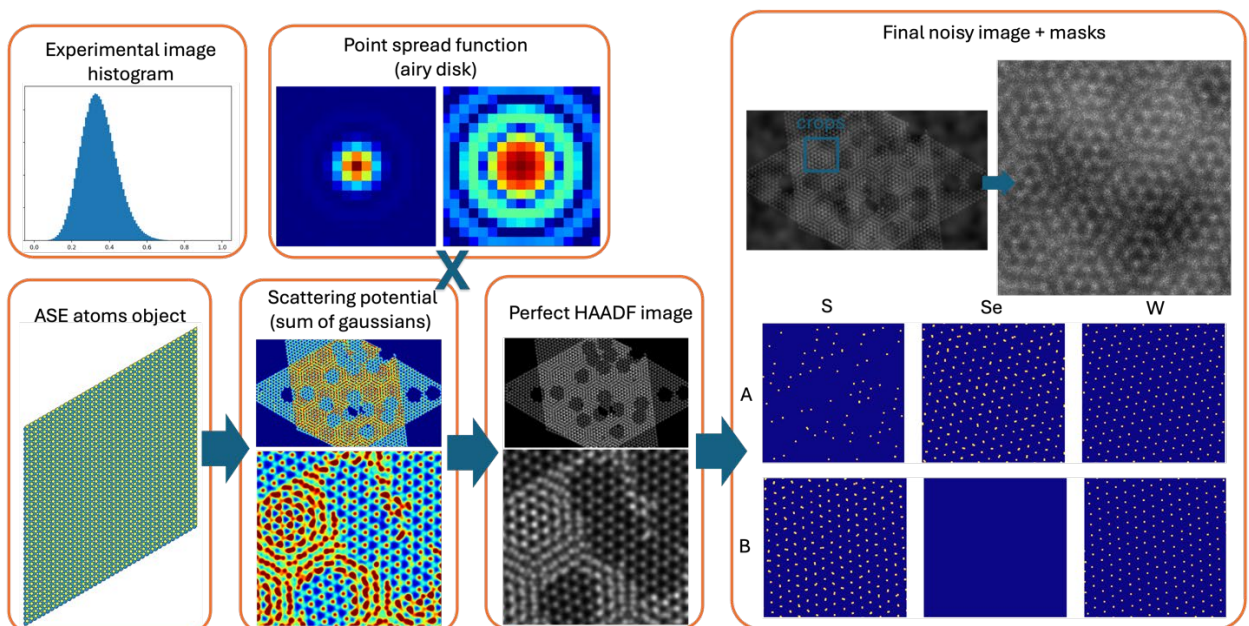

**Figure S1.** An overview of the dataset generation workflow described in the main body of this work, with visualization. Any common atomic coordinates file is converted to an ASE Atoms object in Python. This lattice is replicated and both crystals undergo the transformations listed in Table S1. The lattices are rotated and projected into 2D by adding Gaussians with amplitudes according to the atomic number of the atom. The resulting scattering potential is convoluted with an airy disk kernel, representing the ideal focused electron beam in a system with circular lenses, producing a clean HAADF image. The final step involves adding Poisson (shot) noise and blobby Gaussian noise before cropping the images and ground truth masks (these are calculated from the atom positions. Each cropped image is normalized between 0 and 1.

*Network architecture and training*

The encoder is comprised of sequential convolution blocks, each containing two 2D convolutions with a (3x3) kernel using padding and stride of 1, ReLU[2] activation functions, and a 2D max pooling layer with a (2x2) kernel with stride of 2 for down-sampling. The first encoder block has 32 filters, and each sequential block doubles the number used in the previous block up to a maximum of 256 and 512 for the graphene and TMD models, respectively. Notably, there is a dropout layer in the bottleneck, with dropout set to 0.2 or 0.423 for the graphene and TMD models respectively, as we found this greatly improved generalization. The decoder legs are identical and are the reverse of the encoder, using 2D transpose convolutions for up-sampling. Skip connections are made by concatenating the output of each encoder block with the input of the appropriate decoder block. We used the Adam optimizer[3] for training and tuned the hyperparameters for the TMD model only using Ray Tune[4] with the Optuna search algorithm[5]. The hyperparameters used for all models are available in **Table S2**.

**Network Parameters**

| Parameter | U-Net | U-Net w/ Gomb Loss | Gomb-Net (G) | Gomb-Net (WSSe) |
|---|---|---|---|---|
| N classes | 2 | 2 | 2 | 6 |
| N filters | [32, 64, 128, 256] | [32, 64, 128, 256] | [32, 64, 128, 256] | [32, 64, 128, 256, 512] |
| Dropout | 0.2 | 0.2 | 0.2 | 0.423 |
| Learning rate | 0.00001 | 0.00001 | 0.00001 | 0.001 |
| Loss | Dice Loss | Gomb Loss | Gomb Loss | Gomb Loss |
| epsilon | 1.00E-06 | 1.00E-06 | 1.00E-06 | 1.00E-06 |
| alpha | 2 | 2 | 2 | 1.58 |
| Trainable params | 1927874 | 1927874 | 2682082 | 43235014 |

**Table S2.** Parameters used for initializing network architectures and training the networks. The alpha term is included in the loss function to minimize false positive network outputs. Epsilon is the term for numerical stability in the loss function.

*Loss Function*

Choosing an appropriate loss function is crucial in training neural networks because it measures the discrepancy between predicted outputs and actual targets, guiding the network to adjust its weights to optimize performance for the specific task. For example, the Dice coefficient loss is highly effective for image segmentation tasks due to its ability to measure the overlap between predicted and actual segmentations, making it ideal for applications like medical image analysis[6].

Gomb-Net implements a physics-informed loss function. Because HAADF images are 2D projections of 3D objects, under the kinematic scattering approximation, the atomic ordering in the out-of-plane dimension has no effect on the final image contrast, which is especially true for moiré bilayer systems. Thus, imposing a fixed layer ordering on the network's predictions introduces unnecessary constraints that can hinder its performance. Therefore, a conventional loss function that enforces such an order is not suitable for our task. To address this, we implement a new groupwise combinatorial loss function (Gomb Loss), given by **Eq 1**:

$$TotalLoss = \frac{1}{N}\sum_{i=1}^{N} \frac{\left(\mathrm{DL}(T_{i,A}, T_{i,B}) - \mathrm{DL}(O_{i,A}, O_{i,B})\right)^2}{\left(\mathrm{DL}(O_{i,A}, T_{i,A}) + \mathrm{DL}(O_{i,B}, T_{i,B}) + \varepsilon\right)^{-1} + \left(\mathrm{DL}(O_{i,A}, T_{i,B}) + \mathrm{DL}(O_{i,B}, T_{i,A}) + \varepsilon\right)^{-1} + \varepsilon} \quad (1)$$

where DL(.) is a Dice loss calculated between the targets *T* and the model outputs *O* for class *i* in layer A or B, N is the number of classes, and $\varepsilon$ is a small value to prevent division by zero set to $1\times10^{-6}$. The Dice loss is weighted to penalize false positives. Groupwise refers to the grouping together of the network outputs according to their branch (A or B in this case). This introduces a constraint indicating to the network that different atomic species in the same layer will be strongly correlated to each other. Combinatorial refers to the loss calculation between every combination of these output groups and target groups. This approach disregards the ordering of the outputs, focusing instead on the interlayer coherence and correspondence to the targets.

To achieve this mathematically, two combinatorial loss scenarios are considered. In one scenario, the network output matches the arbitrary mask ordering (left term in the denominator). In the other scenario, the output matches the inverse mask ordering (right term in the denominator). Because only one scenario is true at a time, one loss is small, and the other is large. To weight the scenario with the smaller loss, we take the harmonic mean for the losses of every scenario. Because this function is Schur-concave, its output will always be between the minimum loss and n * the minimum loss (in this case, n = 2). The output of this function is close to the loss of the correct ordering scenario, with a small contribution from the incorrect scenario and the stabilizing constant ε. This small contribution decreases as the network trains.

The loss values between the output and targets are computed with a modified Dice coefficient loss, which penalizes false positive guesses. The purpose of the numerator term is to discourage the network from predicting the same output from each branch, even in cases where the targets for each layer are similar. Before the introduction of this term, the network was predicting the same output for both layers consistently. It is possible that this loss function can be improved, modified, and even applied to problems in other fields that involve a non-specific ordering of grouped network outputs.

*HAADF-STEM Imaging*

HAADF-STEM images of graphene and TMD bilayers were captured on a monochromated, probe corrected Thermo Scientific Spectra 300 operating at 60 kV accelerating voltage with a convergence angle of 30 mrad and a nominal screen current of 100 pA. The collection angle for the samples differed, with 80-200 mrad for the $WS_2$-$WS_{2(1-x)}Se_{2x}$ dataset and 40-200 mrad for the graphene dataset.

*Sample Preparation*

Synthesis of twisted bilayer graphene was prepared following the procedure outlined in Wang et al.[7] This involved the direct transfer of chemical vapor deposition (CVD)-synthesized graphene from Cu foil onto a holey carbon covered TEM grid, followed by precleaning in ambient air at 250 °C. The sample was then exposed to air of near-saturation humidity overnight. Once in the microscope, the sample was further cleaned by a 30-minute beam shower, achieving a level of cleanness that allows a very intense STEM probe to operate continuously for hours without electron-beam induced deposition of hydrocarbons[7]. The bilayer region shown was formed either during the graphene CVD growth or as a result of accidental fold-over during the transfer process.

The twisted $WS_2$ bilayer samples were prepared by stacking two MOCVD grown $WS_2$ monolayers using a wet transfer method as described in Wang et al.[8] This twisted $WS_2$ bilayer was then implanted with Se atoms using a maximum kinetic energy of ~8.5 eV/atom via PLD to form fractional Janus WSSe on the exposed monolayer, resulting in $WS_2$-$WS_{2(1-x)}Se_{2x}$. Full details of the Se implantation can be found in Lin et al.[9] and Harris et al.[10]

*Kernel Density Estimate Bootstrapping*

The bandwidth parameter, was set to 0.1, chosen by Silverman's rule of thumb to balance bias and variance in the density estimates. To quantify the uncertainty in our KDE estimates, we employed a bootstrapping approach. We generated 1000 bootstrap samples by resampling the original dataset with replacement, each containing the same number of data points as the original dataset. For each bootstrap sample, we computed the KDE using the same parameters as described above. The mean and standard deviation of the KDE values were then calculated across the 1000 bootstrap samples.